# Big Data Visualization Tools*

Nikos Bikakis

ATHENA Research Center, Athens, Greece

Data visualization and analytics are nowadays one of the corner-stones of Data Science, turning the abundance of Big Data being produced through modern systems into actionable knowledge. Indeed, the Big Data era has realized the availability of voluminous datasets that are dynamic, noisy and heterogeneous in nature. Transforming a data-curious user into someone who can access and analyze that data is even more burdensome now for a great number of users with little or no support and expertise on the data processing part. Thus, the area of data visualization and analysis has gained great attention recently, calling for joint action from different research areas and communities such as information visualization, data management and mining, human-computer interaction, and computer graphics. This article presents the limitations of traditional visualization systems in the Big Data era. Additionally, it discusses the major prerequisites and challenges that should be addressed by modern visualization systems. Finally, the state-of-the-art methods that have been developed in the context of the Big Data visualization and analytics are presented, considering methods from the Data Management and Mining, Information Visualization and Human-Computer Interaction communities.

## Synonyms

Exploratory data analysis; Information visualization; Interactive visualization; Visual analytics; Visual exploration

## Definition

*Data visualization* is the presentation of data in a pictorial or graphical format, and a *data visualization tool* is the software that generates this presentation. Data visualization offers intuitive ways for information perception and manipulation that essentially amplify the overall cognitive performance of information processing, enabling users to effectively identify interesting patterns, infer correlations and causalities, and support sense-making activities.

*This article appears in: *Encyclopedia of Big Data Technologies*, **2nd Edition**, Springer, 2022
https://doi.org/10.1007/978-3-319-63962-8_109-2

# Overview

Data visualization provides users with intuitive means to interactively explore and analyze data, enabling them to identify interesting patterns, discover correlations and causalities, and support sense-making activities (Throughout the article, terms *visualization* and *visual exploration*, as well as terms *tool* and *system*, are used interchangeably.). This is of great importance, especially given the massive volumes of digital information concerning nearly every aspect of human activity that are currently being produced and collected.

Data visualization and analytics are nowadays one of the cornerstones of Data Science, turning the abundance of Big Data being produced through modern systems into actionable knowledge. Indeed, the Big Data era has realized the availability of voluminous datasets that are dynamic, noisy, and heterogeneous in nature. Transforming a data-curious user into someone who can access and analyze that data is even more burdensome now for a great number of users with little or no support and expertise on the data processing part. Thus, the area of data visualization and analysis has gained great attention recently, calling for joint action from different research areas and communities such as information visualization, data management and mining, human-computer interaction, and computer graphics.

Several traditional problems from those communities, such as efficient data storage, querying and indexing for enabling visual analytics, ways for visual presentation of massive data, efficient interaction, and personalization techniques that can fit to different user needs, are revisited with Big Data in mind (Andrienko et al. 2020; Qin et al. 2020; Idreos et al. 2015; Behrisch et al. 2019; Godfrey et al. 2016; Shneiderman 2008).

Given the above, modern visualization systems should effectively and efficiently handle the following aspects:

- *Real-Time Interaction*. Efficient and scalable techniques should support the interaction with billion-objects datasets while maintaining an acceptable system response in in less than a second.
- *On-the-Fly Visualization*. Support of on-the-fly visualizations over large and dynamic sets of volatile raw (i.e., not preprocessed) data is required. In several cases, a preprocessing phased is not an option.
- *Visual Scalability*. Provision of effective data abstraction mechanisms is necessary for addressing problems related to visual information overloading (aka overplotting).
- *User Assistance and Personalization*. Encouraging user comprehension and offering customization capabilities to different user-defined exploration scenarios and preferences according to the analysis needs are important features.

The literature on visualization is extensive, covering a large range of fields and many decades (Rees and Laramee 2019; McNabb and Laramee 2017). Data visualization is discussed in a great number of recent introductory-level textbooks, such as Ward et al. (2015), Keim et al. (2010). Further, surveys of Big Data visualization systems can be found at (Qin et al. *2020*; Po et al. *2020*; Godfrey et al. *2016*; Behrisch et al. *2019*; Bikakis and Sellis *2016*; Idreos et al. *2015*).

Finally, there is a great deal of information regarding visualization tools available in the Web. We mention dataviz.tools (http://dataviz.tools) and datavizcatalogue (www.datavizcatalogue. com) which are catalogs containing a large number of visualization tools, libraries, and resources.

## Visualization in Big Data Era

This section discusses the basic concepts related to Big Data visualization. First, the limitations of traditional visualization systems are outlined. Then, the basic characteristics of data visualization in the context of Big Data era are presented. Finally, the major prerequisites and challenges that should be addressed by modern visualization systems are discussed.

**Traditional Visualization Systems.** Most *traditional visualization systems* perform well for ad hoc visualizations of small data files (e.g., showing a trend line or a bar chart) or over aggregated data (e.g., summaries of data points, into which user can zoom in), which can fit in main memory. Hence, they restrict themselves to dealing with *small datasets*, which can be easily handled and analyzed with conventional data management and visual explorations techniques. For larger data files, the conventional systems usually require a *preprocessing phase*, such as loading in a data management system. As a result, they are limited to accessing *preprocessed sets of static data*.

**Big Data Era.** On the other hand, nowadays, the Big Data era has made available large numbers of *very big* datasets that are often *dynamic* and characterized by high *variety* and *volatility*. For example, in several cases (e.g., scientific databases), new data constantly arrive on an hourly basis; in other cases, data sources offer query or API endpoints for online access and updating. Further, nowadays, an increasingly large number of *diverse users* (i.e., users with different preferences or skills) explore and analyze data in a plethora of *different scenarios and tasks*.

**Visualization Systems in Big Data Era.** *Modern systems* should be able to efficiently handle *big dynamic datasets*, operating on machines with limited computational and memory resources (e.g., laptops). The dynamic nature of nowadays data (e.g., stream data), hinders the application of a preprocessing phase, such as traditional database loading and indexing. Hence, systems should provide *on-the-fly processing and visualization* over large sets of data.

Further, in conjunction with performance issues, modern systems have to address challenges related to *visual presentation*. Visualizing a large number of data objects is a challenging task; modern systems have to "*squeeze a billion records into a million pixels*" (Shneiderman *2008*). Even in small datasets, offering a dataset overview may be extremely difficult; in both cases, *information overloading* (aka overplotting) is a common issue. Consequently, visual scalability is a basic requirement of modern systems, which have to effectively support data reduction/abstraction (e.g., sampling, aggregation) over enormous numbers of data objects.

Apart from the aforementioned requirements, modern systems must also satisfy the *diversity of preferences and requirements* posed by *different users and tasks*. Modern systems should provide the user with the ability to customize the exploration experience based on her preferences and the individual requirements of each examined task. Additionally, systems should automatically adjust their parameters by taking into account the *environment setting and available resources*, e.g., screen resolution/size, available memory.

## Key Research Findings

This section presents how state-of-the-art approaches from data management and mining, information visualization, and human-computer interaction communities attempt to handle the challenges that arise in the Big Data era.

**Data Reduction.** In order to handle and visualize large datasets, modern systems have to deal with information overloading issues. Offering *visual scalability* is crucial in Big Data visualization. Systems should provide efficient and effective abstraction and summarization mechanisms. In this direction, a large number of systems use *approximation techniques* (aka data reduction techniques), in which abstract sets of data are computed. Considering the existing approaches, most of them are based on (1) *sampling* and *filtering* (Fisher et al. *2012*; Park et al. *2016*; Agarwal et al. *2013*; Battle et al. *2013*) and/or (2) *aggregation* (e.g., binning, clustering) (Elmqvist and Fekete *2010*; Bikakis et al. *2017*; Jugel et al. *2015*; Liu et al. *2013*).

**Hierarchical Data Exploration.** Data reduction techniques are often defined hierarchically (Elmqvist and Fekete *2010*), allowing users to explore data in multiple "level of detail" by, e.g., hierarchical aggregation.

*Hierarchical approaches* (aka multilevel) allow the visual exploration of very large datasets in multiple levels (with different "level of detail"), offering both an overview and an intuitive and effective way for finding specific parts within a dataset.

Particularly, in hierarchical approaches, the user first obtains an overview of the dataset before proceeding to data exploration operations (e.g., Roll-Up, Drill-Down, Zoom, Filter) and finally retrieving details about the data. A significant challenge, in large data visualization, is the problem of overplotting. It can effectively be addressed in hierarchical approaches, in which, in each level, the number of the presented visual elements is controlled by data reduction methods.

Hierarchical techniques have been extensively used in large graphs/network visualization, in order to handle the common problem of overloading, in winch the graph is presented as "hairball." In these techniques the graph is recursively decomposed into smaller subgraphs that form a hierarchy of abstraction layers. In most cases, the hierarchy is constructed by exploiting clustering and partitioning (Rodrigues Jr. et al. *2013*; Tominski et al. *2009*), sampling (Sundara et al. *2010*), and edge bundling (Gansner et al. *2011*) techniques.

**Progressive Data Visualization.** Visual data exploration requires real-time system's response. However, computing complete results over large (unprocessed) datasets may be extremely costly and in several cases unnecessary. Modern systems should progressively return partial and preferably representative results, as soon as possible (Angelini et al. 2018; Zgraggen et al. 2017).

*Progressiveness* can significantly improve efficiency in exploration scenarios, where it is common that users attempt to find something interesting without knowing what exactly they are searching for beforehand. In this case, users perform a sequence of operations (e.g., queries), where the result of each operation determines the formulation of the next operation.

Recently, many systems adopt the *progressive paradigm* attempting to reduce the response time (Moritz et al. 2017; Rahman et al. 2017; Angelini et al. 2018; Zgraggen et al. 2017; Fisher et al. 2012; Agarwal et al. 2013). Progressive approaches, instead of performing all the computations in one step (that can take a long time to complete), split them in a series of short chunks of approximate computations that improved with time. Therefore, instead of waiting for an unbounded amount of time, users can see the results unfolding progressively. This way the users are able to interrupt the execution and define the next operation, without waiting the exact result to be computed.

**Adaptive Indexing and In-situ Data Management.** Several approaches like database cracking and adaptive indexing have been adopted in data exploration scenarios. The basic idea of these is to incrementally adapt the indexes and/or refine the physical order of data, during query processing, following the characteristics of the workload (Pedro et al. 2019; Vikram et al. 2020; Matheus et al. 2021; Stratos et al. 2007).

In-situ paradigm (Idreos et al. 2011; Alagiannis et al. 2012; Bikakis et al. 2021; Maroulis et al. 2022; Olma et al. 2017) is a recent trend that aims at enabling the on-the-fly querying over large sets of raw data, by avoiding the (pre)processing (e.g., loading and indexing) overhead of traditional DBMS techniques. In-situ query processing aims at avoiding data loading in a DBMS by accessing and operating directly over raw data files. In these systems, in situ *incremental and adaptive* processing and indexing techniques are used, in which small parts of raw data are processed incrementally "following" users' interactions.

Furthermore, several well-known DBMS support in situ SQL querying over CSV files. Particularly, MySQL provides the CSV Storage Engine, Oracle offers the External Tables and Postgres has the Foreign Data.

**Visual-Oriented Data Structures.** Numerous data visualization systems have been developed on-top of data structures and indexes which are designed in the context of visual exploration, to improve efficiency and scalability. VisTrees (El-Hindi et al. 2016) and HETree (Bikakis et al. 2017) are tree-based main-memory indexes that address visual exploration use cases; i.e., they offer exploration-oriented features such as incremental index construction and adaptation.

Nanocubes (Lins et al. 2013), Hashedcubes (de Lara Pahins et al. 2017), SmartCube (Liu et al. 2020), Gaussian Cubes (Wang et al. 2017), and TopKubes (Miranda et al. 2017) are main-memory data structures supporting interactive visualization. They are based on main-memory variations of a data cube in order to reduce the time needed to generate the visualization.

Spatial indexes have been used in graphVizdb (Bikakis et al. 2016) is a graph-based visualization tool, which employs a 2D spatial index (e.g., R-tree) and maps user interactions into window 2D queries. Spatial 2D indexing is also adopted in Kyrix (Wenbo et al. 2019), enabling efficient Zoom and Pan operations over arbitrary data types. Finally, tile-based structures are used in several visualization systems, such as RawVis (Bikakis et al. *2021*), AID (Saheli et al. *2019*), ForeCache (Battle et al. *2016*).

**Caching and Prefetching.** Recall that, in exploration scenarios, a sequence of operations is performed and, in most cases, each operation is driven by the previous one. In this setting, *caching* and/or *prefetching* the sets of data that are likely to be accessed by the user in the near future can significantly reduce the response time (Battle et al. *2016*; Tauheed et al. *2012*). Most of these approaches use prediction techniques which exploit several factors (e.g., user behavior, user profile, use case) in order to determine the upcoming user interactions.

**User Assistance.** The huge amount of available information makes it difficult for users to manually explore and analyze data. Modern systems should provide mechanisms that assist the user and reduce the effort needed on their part, considering the diversity of preferences and requirements posed by different users and tasks.

Recently, several approaches have been developed in the context of *visualization recommendation* (Vartak et al. *2016*). These approaches recommend the most suitable visualizations in order to assist users throughout the analysis process. Usually, the recommendations take into account several factors, such as data characteristics, environment setting and available resources (e.g., screen resolution/size, available memory), examined task, user preferences and behavior, etc.

Considering data characteristics, there are several systems that recommend the most suitable visualization technique (and parameters) based on the type, attributes, distribution, or cardinality of the input data (Key et al. *2012*; Ehsan et al. *2016*). Other approaches provide visualization recommendations based on user behavior and preferences (Mutlu et al. *2016*), using machine learning (Hu et al. *2019*) or similarity-based techniques (Kim et al. *2017*). In a similar context, some systems assist users by recommending certain visualizations that reveal surprising, interesting data or outliers (Vartak et al. *2014*; Wongsuphasawat et al. *2016*).

## Examples of Applications

Visualization techniques are of great importance in a wide range of application areas in the Big Data era. The volume, velocity, heterogeneity, and complexity of available data make it extremely difficult for humans to explore and analyze data. Data visualization enables users to perform a series of analysis tasks that are not always possible with common data analysis techniques (Keim et al. *2010*).

Major application domains for data visualization and analytics are *Physics* and *Astronomy*. Satellites and telescopes collect daily massive and dynamic streams of data. Using traditional analysis techniques, astronomers are able to identify noise, patterns, and similarities. On the other hand, visual analytics can enable astronomers to identify unexpected phenomena and perform several complex operations, which are not are feasible by traditional analysis approaches.

Another application domain is *atmospheric sciences* like *Meteorology* and *Climatology*. In this domain high volumes of data are collected from sensors and satellites on a daily basis. Storing these data over the years results in massive amounts of data that have to be analyzed. Visual analytics can assist scientists to perform core tasks, such as climate factors correlation analysis, event prediction, etc. Further, in this domain, visualization systems are used in several scenarios in order capture real-time phenomena, such as hurricanes, fires, floods, and tsunamis.

In the domain of *Bioinformatics*, visualization techniques are exploited in numerous tasks. For example, analyzing the large amounts of biological data produced by DNA sequencers is extremely challenging. Visual techniques can help biologist to gain insight and identify interesting "areas" of genes on which to perform their experiments.

In the Big Data era, visualization techniques are extensively used in the *business intelligence* domain. *Finance markets* is one application area, where visual analytics allow to monitor markets, identify trends, and perform predictions. Besides, *market research* is also an application area. Marketing agencies and in-house marketing departments analyze a plethora of diverse sources (e.g., finance data, customer behavior, social media). Visual techniques are exploited to realize task such as identifying trends, finding emerging market opportunities, finding influential users and communities, and optimizing operations (e.g., troubleshooting of products and services), business analysis, and development (e.g., churn rate prediction, marketing optimization).

## Future Directions for Research

From a community perspective, the challenges related to data visualization and analysis involve different communities and research areas such as Data management & Mining, Information Visualization, Human-Computer Interaction, and Computer Graphics. In what follows we summarize some of the basic challenges indicated in a recent report (Andrienko et al. *2020*).

**Understand needs, personalize, and guide.** Modern systems need to handle several major user-centric challenges. Systems should understand what the users need to solve their problems and offer guidance ("Show the Data not Seen by Humans"). In this context, the following basic challenge can be considered: (a) recommend views of the data that the users might want to analyze; (b) find what parts of data will be useful for each task; (c) provide insights recommendations; (d) produce data stories and explanations; (e) develop novel interfaces that assist users to understand data types and properties of the data; (f) integrated human factors related to human vision and perception to analysis pipeline, so users supervise, or provide feedback to systems.

**Scalability and efficiency.** Another great challenge is related to the systems' scalability and efficiency. This is to enable visualization systems to efficiently handle billion objects datasets, while limiting the response to a few milliseconds. In that direction, the challenges involve how to build tools that can perform interactive operations and complex analytics over massive sets of data. In that respect, there is the need for novel approaches (e.g., progressive data processing) that can handle large streaming, sampled, uncertain, high-dimensional, and noisy data.

**Data-intensive applications.** Classical data management problems, such as data storage, querying, and indexing, are highly related to efficiency and scalability of the modern visualization systems. However, in the context of visual analysis, solving such problems reveals several "new" challenges. Such challenges are considered the following: define visualization-centric algebras, design visualization operators, implement operation optimization techniques, define effective storage and indexing scheme.

**Interactive machine learning.** Building interactive tools and enabling visual analysis to Machine Learning (ML) applications is a great challenge. For example, develop visual methods for interpreting and techniques for interacting with ML models; implement visualization systems that enable models' troubleshooting, debugging, and comparison.

## Cross-References

- Visualization
- Visualization Techniques
- Visualizing Semantic Data
- Graph Exploration and Search